\documentclass[journal=jacsat,manuscript=letter]{achemso}

\usepackage{xcolor}
\usepackage[utf8]{inputenc}
\usepackage{amsmath}
\usepackage{amssymb}
\usepackage{graphicx}
\usepackage{gensymb}
\usepackage{graphics}
\usepackage[T1]{fontenc}
\usepackage[]{epstopdf}

\author{Ludovica Zullo}
\affiliation{Department of Physics, University of Trento, Via Sommarive 14, 38123 Povo, Italy}
\alsoaffiliation{Sorbonne Universit\'e, CNRS, Institut des Nanosciences de Paris, UMR7588, F-75252 Paris, France}
\email{ludovica.zullo@unitn.it}
\author{Giovanni Marini}
\affiliation{Graphene Labs, Fondazione Istituto Italiano di Tecnologia, Via Morego, I-16163 Genova, Italy}
\author{Tristan Cren}
\affiliation{Sorbonne Universit\'e, CNRS, Institut des Nanosciences de Paris, UMR7588, F-75252 Paris, France}
\author{Matteo Calandra} 
\affiliation{Department of Physics, University of Trento, Via Sommarive 14, 38123 Povo, Italy}
\alsoaffiliation{Sorbonne Universit\'e, CNRS, Institut des Nanosciences de Paris, UMR7588, F-75252 Paris, France}
\alsoaffiliation{Graphene Labs, Fondazione Istituto Italiano di Tecnologia, Via Morego, I-16163 Genova, Italy}
\email{m.calandrabuonaura@unitn.it}

\title{Misfit layer compounds as ultra-tunable field effect transistors: from charge transfer control to emergent superconductivity}
\begin{document}

\begin{abstract}
Misfit layer compounds are heterostructures composed of rocksalt units stacked with few layers transition metal dichalcogenides. They host Ising superconductivity, charge density waves and good thermoelectricity. The design of misfits emergent properties is, however, hindered by the lack of a global understanding of the electronic transfer among the constituents. Here, by performing  first principles calculations, we unveil the mechanism controlling the charge transfer and demonstrate that rocksalt units are always donor and dichalcogenides acceptors. We show that misfits behave as a periodic arrangement of ultra-tunable field effect transistors where a charging as large as $\approx6\times10^{14}$ e$^-$cm$^{-2}$  can be reached and  controlled efficiently by the La-Pb alloying in the rocksalt. Finally, we identify a strategy to design emergent superconductivity and demonstrate its applicability in (LaSe)$_{1.27}$(SnSe$_2$)$_2$. Our work paves the way to the design synthesis of misfit compounds with tailored physical properties.
\end{abstract}

The capability of inducing a controlled and tunable number of carriers in few layer systems has been pivotal for the success of 2D materials~\cite{Wu2023}. However, in metallic few layers 2D dichalcogenides such as NbSe$_2$, the largest carrier doping that can be achieved via field effect gating are of the order of $n_e\approx 3\times 10^{14}$ e$^-$ cm$^{-2}$~\cite{PhysRevLett.117.106801}, corresponding  to a Fermi level shift of the order of $0.1$ eV, too small to drastically change the physical properties. 

Recently ~\cite{MisfitsMCTC2021}, it has been shown that overcoming this limit is possible in the misfit layer compound (MLC) (LaSe)$_{1.14}$ (NbSe$_{2}$)$_{2}$, an heterostructure composed of periodically alternating rocksalt monocalchogenide units (RS) and few layers transition metal dichalcogenides (TMDs)~\cite{WIEGERS19961,WIEGERS_ROUXEL_MLC_1995}. In this system, a massive electron transfer from the LaSe RS to the NbSe$_2$ TMD occurs, leading to a rigid Fermi level shift as large as $+0.55$ eV. It is, however, unclear if the electron doping in misfits can be in some way controlled by any physical parameter and, more important, how general this mechanism to dope few layer TMDs is.

MLCs have been known for a long time and their structures as a function of the RS and TMD composition have been thoroughly investigated ~\cite{WIEGERS19961,WIEGERS_ROUXEL_MLC_1995}. However, the exploration of physical properties such as Ising superconductivity~\cite{TristanSupercondMLCs,BOBCAVA_MLCsSupercond,SUPERCOND_MISFITS_KIM20211,Sofranko2020,SUPERCOND_MISFITS_Yang_2019,SupercondGrosse2016} charge density waves (CDW) ~\cite{CDW_MISFIT_ATKINS2013,BiSe_TiSe2_NO_CDW_TRUMP2014,CDW_MISFIT_FALMBIGL2015,CDW_MISFITS_2022,ZHU_2023_MISFITS} or topological effects ~\cite{TOPOL_MISFITS_CAVA2016} are quite recent. The research in the field has lead to remarkable results but it has mostly proceeded by isolated discoveries and trial and error chemical synthesis, while general rules to understand what happens when assembling different RS and TMDs are missing. The need of a global picture becomes evident when considering that (i) many ternary alloys composed of monochalcogenides can be assembled with practically any few layer dichalchogenide, (ii) the thickness of the dichalcogenide layers can be chosen at will. This makes a lot of possible combinations and leads to many unanswered questions. 
For example, how does the charge transfer occur in these structures ? Are the TMD layers  acceptors or donors ? How can the charge transfer be tuned ? To what extent the electronic structure of the TMD is affected when inserted in the heterostructure ? Most important, what are the emergent properties of the misfit, i.e. properties of the MLC that are absent in the pristine constituents ? How can we design misfit properties from the knowledge of their building blocks ?

In this work we answer these questions by performing extensive first principles electronic structure calculations of MLCs. We identify the fundamental mechanism ruling charge transfer and demonstrate how the charge injection into the TMD layers can be efficiently controlled by chemical alloying in the rocksalt unit. Most important, we show that superconductivity can emerge in 
MLCs formed by assembling non-superconducting RS and TMDs. Finally, we demonstrate that misfit layer compounds can be assimilated to ultra-tunable field effect transistor with an unequaled charging of the TMD layers. Our work paves the way to extensive experimental synthesis and development of these promising systems.

\begin{figure}[t!]
\centering
\includegraphics[width=1\linewidth]{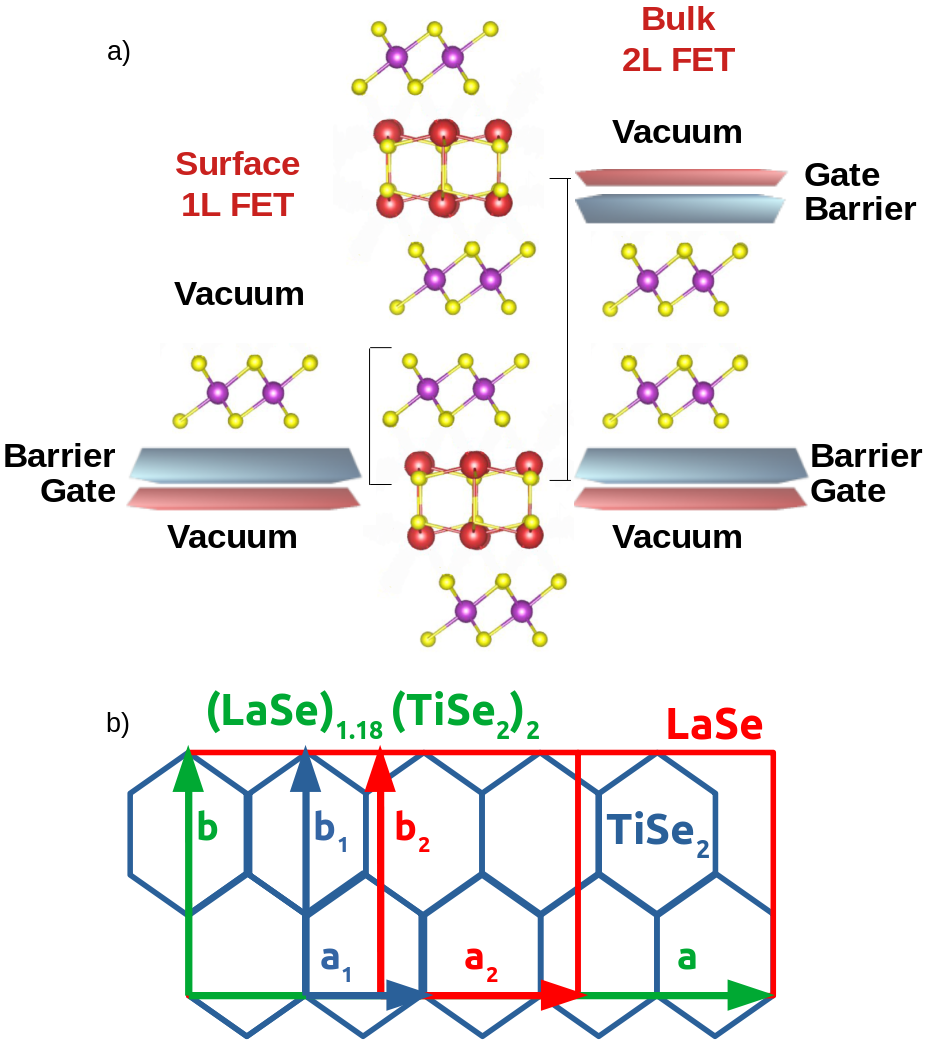}
\caption{a) Bulk structure of (LaSe)$_{1.18}$(TiSe$_{2}$)$_{2}$. The field-effect modeling scheme is depicted for the case of the most common TiSe$_2$ terminated surface (left) and for bulk  TiSe$_{2}$ (right). b) Sketch of the unit cell of (LaSe)$_{1.18}$(TiSe$_{2}$)$_{2}$ misfit layer compound $({\bf a},{\bf b})$ compared with the ones of a TiSe$_2$ monolayer $({\bf a}_1,{\bf b}_1)$ and of a LaSe unit $({\bf a}_2,{\bf b}_2)$.}\label{fig:1}
\end{figure}

The chemical formula of MLCs is (RQ)$_{1+\delta}$(TX$_{2}$)$_{m}$, where (TX$_{2}$)$_{m}$ is a $m-$layers TMD and RQ is a rocksalt monochalcogenide unit (often referred to as Q-layer)~\cite{WIEGERS19961,WIEGERS_ROUXEL_MLC_1995}. Ternary alloys of two monochalcogenides within a single RS Q-layer (e.g. La$_{x}$Sr$_{1-x}$S) have also been synthesized ~\cite{CarioPhysRevB.55.9409} leading to MLCs having chemical formulas of the kind (R$_{x}$M$_{1-x}$Q)$_{1+\delta}$(TX$_{2}$)$_{m}$. As a prototypical example of the MLCs crystal structure we consider (LaSe)$_{1.18}$(TiSe$_{2}$)$_{2}$, shown in Fig. \ref{fig:1} (a) and (b). Each TiSe$_2$ and LaSe sublattice has its
own set of cell parameters. Compared to bulk 1TTiSe$_2$, the lattice of the TiSe$_2$ bilayer in the MLC is not perfectly hexagonal and is slightly expanded along one direction. As a
consequence, the TiSe$_2$ sublattice is described by a centered
orthorhombic cell with in-plane lattice vectors ${\bf a}_1 \approx 3.6$~\AA~  and
${\bf b}_1 \approx 6$~\AA . The LaSe sublattice has also an orthorhombic symmetry but with similar in-plane lattice parameters ${\bf a}_2 \approx {\bf b}_2 \approx 6$~\AA. Both systems have the same ${\bf b}$ vectors (${\bf b}_1\approx {\bf b}_2$)  so that the material is commensurate along this direction. The ratio between the norms of the ${\bf a}_1$ and ${\bf a}_2$ vectors sharing the same direction is an irrational number (see tables in Fig. 1 and 2 in Supplemental Material) making the MLC incommensurate in the ${\bf a}$ direction. The mismatch ratio $a_{2}/a_{1}=x/y$ is usually in the range $\sim 1.6-1.8$ and sets the parameter $\delta$ in the chemical formula through the relation $1+\delta=2\times (a_{1}/a_{2})$. 
In this work we adopt the convention of using the value of $\delta$ as obtained from the lattice parameters {\bf a}$_{1}$ and {\bf a}$_{2}$  of the pristine RS and TMD before assembling them in a MLC structure, as reported in the tables in Figs. 1-2 in the Supplemental Material. The commensurate approximant of each MLCs considered in the current work is reported in Fig. 3 in the Supplemental Material.

\begin{figure}[t!]
\centering
\includegraphics[width=1\linewidth]{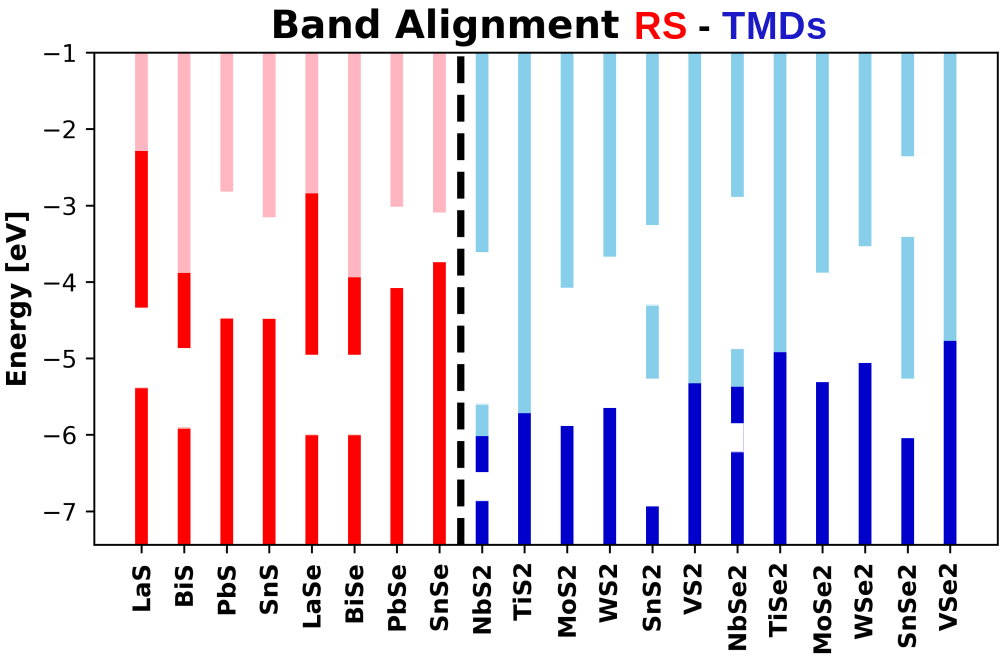}
\caption{Calculated band-alignement of isolated Q-layer rocksalt structures (red) and single layer transition metal dichalcogenides (blue). Dark (light) bars represents the position of E$_{F}$/valence band maximum (E$_{F}$/conduction band minimum) for metal/insulators. White spaces in the  bars represent gaps in the single particle spectrum. The zero of the energy is set to the vacuum level.}\label{fig:2}
\end{figure}

The RS layers have a strong intralayer  bonding. A strong bonding also forms among the RS and TMDs layers. On the contrary, Van der Waals bonding occurs among the closer TMD layers. After cleavage, for  $m>1$, the surface of the sample is a perfect TMD layer (a single layer in the $m=2$ case considered in this work ~\cite{MisfitsMCTC2021}).
In the $m=1$ case , i.e. a single layer TMD sandwiched among RS Q-layers, the bonding along the $z$ axis is always strong. As a result, the cleavage occurs in-between the RS and TMD bonding and the surface is still a TMD single layer, however it is often less clean and presents several steps and defects~\cite{TristanSupercondMLCs, Sofranko2020}. In all cases, there is a substantial experimental evidence ~\cite{MisfitsMCTC2021} that ARPES and STS/STM measurements mostly sample the terminating TMD layer without accessing the bulk of the structure. On the contrary, Raman, transport and superconducting measurements probe bulk properties of the crystal.

\begin{figure*}[t!]
\centering
\includegraphics[width=1\linewidth]{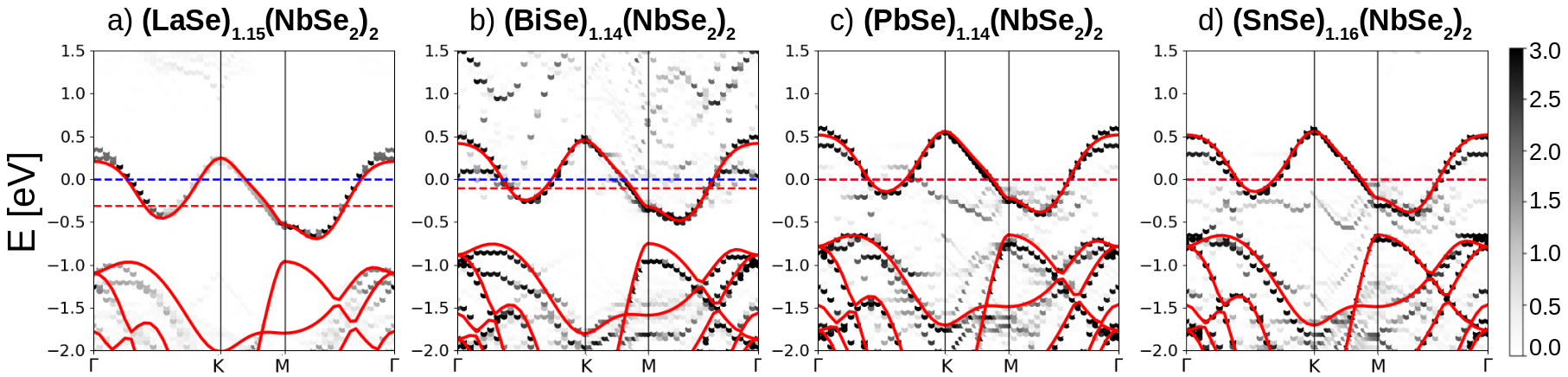}
\caption{Band unfolding onto the NbSe$_2$ single layer Brillouin zone for the NbSe$_2$ misfit series for different rocksalt Q-layers having comparable mismatching ratio close to $7/4$. The band structure for the isolated single layer NbSe$_{2}$ (red line) is superimposed and aligned to the Nb d-band in the misfit. The blue dashed line corresponds to the Fermi level E$_{F}$ of the misfit compound, while the red one to the Fermi level of the isolated NbSe$_2$ layer.  In the last two panels the dashed red line is superimposed to the dashed blue one. Spin orbit coupling is neglected in all the calculations.}\label{fig:3}
\end{figure*}

In order to gain insight on the charge transfer among the RS and TMD layers in the MLC and its relevance for the electronic structure measurements (ARPES), we perform extensive calculations of the work functions of $8$ isolated rocksalt Q-layers and $12$ isolated TMDs single layers. The choice of considering TMD single layers is motivated by (i) the fact that we consider MLC with $m=2$ having a single layer TMD as terminating surface and (ii) by the fact that the work functions of bilayers TMDs is fairly close to the one of single layers ~\cite{Band_Alignment_Kim_2021}. Thus, we expect that our results will also hold for the surface and the bulk and for the $m=1$ case. Calculations are performed with the \textsc{quantum ESPRESSO}~\cite{QE} package and we use the PBE exchange and correlation functional ~\cite{PBE} (see SI for more technical details). Results are shown in Fig. \ref{fig:2}.

The key quantities ruling the charge transfer in these systems are the work function difference among RS and TMDs and the consequent band alignment, the lattice mismatching ratio a$_{2}/$a$_{1}$ and, finally,
the degree of hybridization when the two subsystems are in contact.
As shown in Fig. \ref{fig:2}, the TMDs globally possess substantially larger work functions than the RS compounds. As the work function is  the energy required to transfer an electron from the Fermi level  to the vacuum level,  RS are always donor and TMDs always acceptors. The net amount of charge transfer depends, however, not only on the work function difference but also on the mutual concentration of the RS and TMD that is related to the mismatching ratio. To explain this more clearly, each RS can transfer a given amount of charge to the TMDs layer, if the mismatching ratio is close to one. However, if the mismatching ratio increases, the relative concentration of RS atoms per TMD cell decreases, and so does the charge transfer. By looking at Fig. 3 in SI, it is clear that the mismatching ratio varies mostly due to the change in the TMD lattice parameter.

In order to demonstrate this global picture we perform explicit calculations for several misfit surfaces terminated by a single layer NbSe$_2$ but having different RS Q-layers as building blocks and sharing comparable mismatching ratios very close to $7/4$ (these compounds all belongs to the ninth column in the table in Fig. 3 in the Supplemental Material).
As it can be seen in Fig. \ref{fig:3}, the behaviour of the (LaSe)$_{1.15}$ (NbSe$_2$)$_2$, (BiSe)$_{1.14}$(NbSe$_2$)$_2$, (PbSe)$_{1.14}$(NbSe$_2$)$_2$ and (SnSe)$_{1.16}$(NbSe$_2$)$_2$ serie is almost completely characterized by the work function differences. Indeed as W(LaSe)$<$W(SnSe)$<$W(PbSe), the charge transfer decreases by progressively decreasing the difference W(NbSe$_2$)-W(RS), as expected. The work function of BiSe is slightly larger than the one of SnSe, however BiSe seems to transfer few more electrons than SnSe. 
We attribute this to the metallic character of BiSe and the consequent stronger hybridization occurring between BiSe and NbSe$_2$, resulting in a substantial band deformation of the pristine NbSe$_2$, as shown in Fig. \ref{fig:3}.

\begin{figure}[htpb!]
\centering
\includegraphics[width=0.7\linewidth]{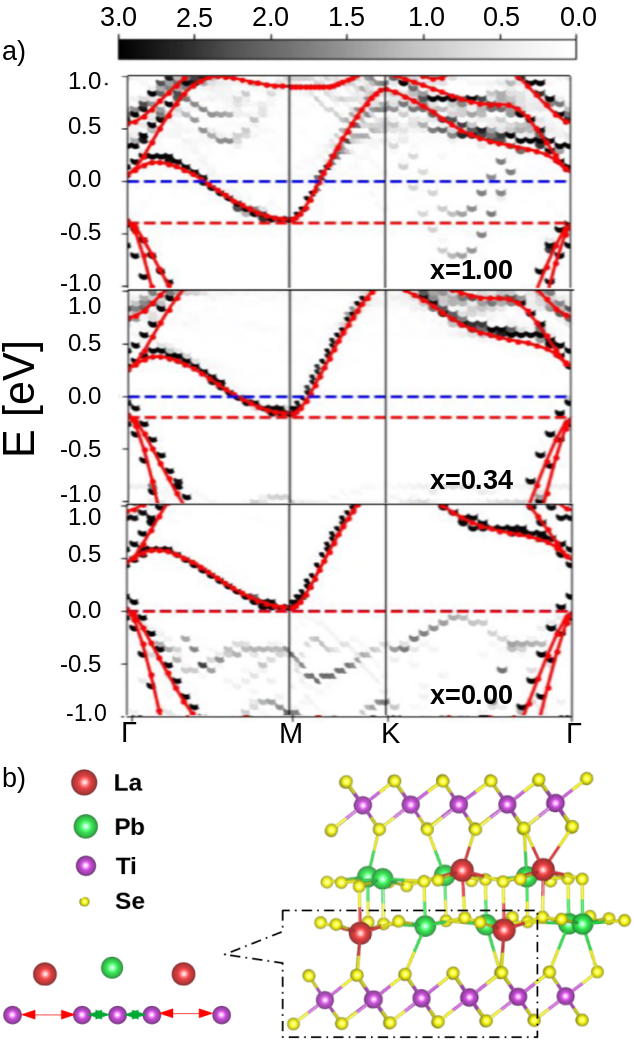}
\caption{a) Band unfolding onto the single layer TiSe$_{2}$ Brillouin zone of the misfit compound (La$_{x}$Pb$_{1-x}$Se)$_{1.18}$(TiSe$_{2}$)$_{2}$ for $x=1.0,0.34,0.0$. The band structure for the isolated single layer TiSe$_{2}$ (red line) is superimposed and aligned to the bottom of the Ti d-band in the misfit. The blue dashed line corresponds to the Fermi level E$_{F}$ of the misfit compound, while the red one to the Fermi level of the isolated TiSe$_2$ layer (in the lowest panel they coincide). b) Lattice deformation of the TiSe$_{2}$ layers generated by the partial substitution of Pb atoms in (La$_{x}$Pb$_{1-x}$Se)$_{1.18}$(TiSe$_{2}$)$_{2}$. The magnified portion shows a bond length alternation in the TiSe$_{2}$ lattice with two different distances $d_{1}$ (red) and $d_{2}$ (green).}\label{fig:4}
\end{figure}

Finally we point out that the NbSe$_2$ electronic structure in going from (PbSe)$_{1.14}$(NbSe$_2$)$_2$ to (LaSe)$_{1.15}$ (NbSe$_2$)$_2$ is n-doped rigidly, i.e. the charge transfer simply induces a Fermi level upshift. From this analysis two questions arise: how general is this rigid doping effect and how can it be used to effectively tune the doping ? We now show that it is possible to engineer the misfit in such a way that the  doping level is rigidly adjustable through appropriate alloying of the RS Q-layer.

For this reason we consider MLCs having the following stoichiometry
(La$_{x}$Pb$_{1-x}$Se)$_{1.18}$ (TiSe$_2$) as a function of $x$.  We point out that similar substitutions (La$\leftrightarrow$Sr) have already been achieved in sulfur-based MLC ~\cite{CarioPhysRevB.55.9409}. A comparison between this system and the previous results for the NbSe$_2$ series will allow us to draw conclusions that are less dependent on the chosen TMD. 

From the previous reasoning and from Fig. \ref{fig:2},  we expect that the La concentration ($x$) allows to tune the carrier concentration in the TiSe$_2$ layers with $x=1$ ($x=0$) corresponding to the highest (lowest) n-doping. 
In Fig.[\ref{fig:4}] we show the calculated band structure of the full (La$_{x}$Pb$_{1-x}$Se)$_{1.18}$(TiSe$_{2}$)$_{2}$ misfit for $x=1.0,0.34,0.0$. We also plot (red continuous line) the electronic structure for an isolated single layer. The position of the bottom of the Ti d-band of the isolated single layer is aligned to the corresponding band in the misfit. As it can be seen, by increasing $x$ the doping is increased. Most important, the Ti d-band displays no deformation upon doping. At the highest doping level ($x=1$, corresponding to a charge transfer of $0.53$  electrons per Ti, which is $n_e\sim 5 \times 10^{14}$ e$^-$ cm$^{-2}$) two parabolic La bands cross the Fermi level along the $\Gamma$K direction. These bands disappear by decreasing $x$ (see SI for calculation at additional values of $x$). Remarkably, the electronic structure of (PbSe)$_{1.18}$(TiSe$_2$)$_2$ is almost indistinguishable from the one of the isolated TiSe$_2$ layer.

Despite this similarity in the electronic structure, we find that (PbSe)$_{1.18}$(TiSe$_2$)$_2$ does not display a $2\times 2$ CDW as it happens in the case of the supported TiSe$_2$ single layer~\cite{2DMaterials-2018-substrate-dependent-TC,Wang2018-advmat,FangPhysRevB.95.201409}. This result is in agreement with resistivity data on this MLC~\cite{BOBCAVA_MLCsSupercond} where no CDW was detected. We attribute the suppression of the CDW to the strong bonding between TiSe$_2$ and the the RS Q-layer. We find that in (La$_{x}$Pb$_{1-x}$Se)$_{1.18}$(TiSe$_{2}$)$_{2}$, for $x\ne0,1$, the Ti-Ti distances are modulated by the presence of Pb atoms in the host LaSe lattice (i.e. the Ti-Ti distance becomes shorter if the Ti atoms are close to a Pb atom).
 The reason is mostly sterical as the La atomic radius  is  larger than the one of Pb, therefore Pb atoms are more strongly bounded to the RS layer and a consequent deformation of the LaSe rocksalt host occurs (as shown in Fig. \ref{fig:4} (b)) followed by a modulation of the Ti-Ti distances. We verified that even starting from $2\times 2$ distorted TiSe$_2$ layers in the  misfit, the structural optimization suppresses the CDW and leads to other distortion patterns that essentially follow the Pb atoms superstructure. Our analysis shows that altering the chemical composition of the rocksalt has a double effect: on the one hand, it allows to precisely tune the rigid doping of the TMD, on the other hand it suppresses the $2\times 2$ CDW of the TiSe$_2$ bilayer and introduces an additional modulation related to the alternation of La and Pb.     

After achieving a complete knowledge of the charge transfer in MLC, we now demonstrate how to design a misfit superconductor starting from its constituents. In particular we show that non-superconducting pristine RS and TMD compounds can lead to a superconductor via charge transfer control (emergent superconductivity).

\begin{figure}[htpb!]
\centering
\includegraphics[width=1\linewidth]{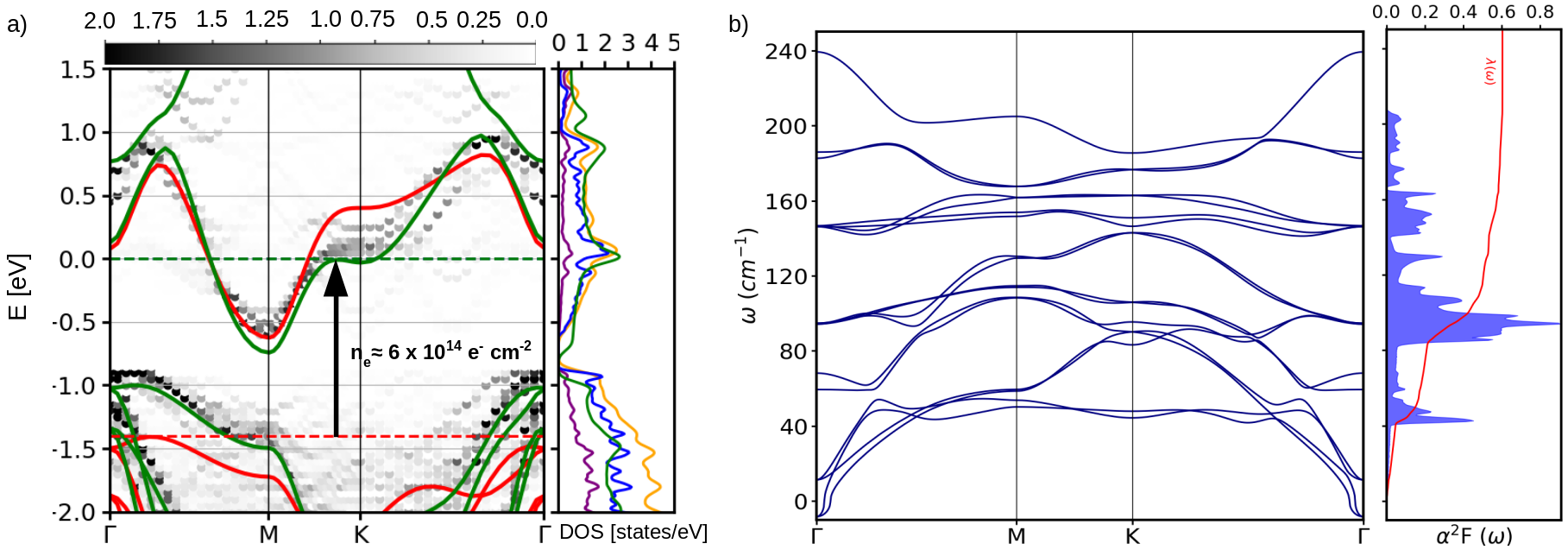}
\caption{Panel a): band unfolding of (LaSe)$_{1.27}$(SnSe$_{2}$)$_{2}$  misfit supercell onto the hexagonal primitive Brillouin Zone (BZ) of single layer SnSe$_{2}$ (the zero  energy is set to the Fermi level, dashed green line) . The superimposed solid lines are the band structure of an isolated single layer $SnSe_{2}$ (red), and of a single layer SnSe$_{2}$ doped in a single FET setup as in (LaSe)$_{1.27}$(SnSe$_{2}$)$_{2}$ by $0.7$ electrons per Sn atoms (green), respectively. Darker regions in the colormap represent the most relevant projection of the misfit eigenvalues of the band structure in the SnSe$_{2}$ first BZ (band unfolding). 
In the adiacent panel we plot the total DOS per SnSe$_{2}$ formula unit of (LaSe)$_{1.27}$(SnSe$_{2}$)$_{2}$ (yellow) and the projected density of states over atomic orbitals of the LaSe layers (purple) and of the SnSe$_2$ layers (blue), respectively. The green line is the DOS of a  single layer SnSe$_{2}$ doped in a single FET setup  of $0.7$ electrons per Sn atoms.
Panel b): dynamical properties and electron-phonon coupling of (LaSe)$_{1.27}$(SnSe$_{2}$)$_{2}$ modeled by a bilayer SnSe$_{2}$ in a double FET setup. The phonon dispersion is shown in the first panel while, in the adjacent panel, the Eliashberg function $\alpha^{2}F(\omega)$ (filled blue curve) and the total electron-phonon coupling $\lambda(\omega)$ (red) are depicted.}\label{fig:5}
\end{figure}

We consider the layered indirect gap semiconductor 1TSnSe$_2$ that can be exfoliated and synthesized in single layer form ~\cite{Advanced_Materials_Interfaces_Fu_2022}. The electronic structure of a single layer SnSe$_2$ is shown in Fig. \ref{fig:5} (red line). The conduction band is formed by an isolated band with a Van Hove singularity point at K. A maximum in the density of states 
occurs at the energy corresponding to the band flattening. If the Fermi level is tuned at the inflection point, this would be beneficial for superconductivity. However, this involves a $\approx 1.4$ eV Fermi level shift corresponding to a charge transfer of $0.77$ electrons ($\approx 6\times10^{14} $e$^-$cm$^{-2}$), unreachable even in a ionic-liquid based field effect transistor. However, as previously shown, this electron doping level could be reached in the misfit (La$_x$Pb$_{1-x}$Se)$_{1.27}$(SnSe$_2$)$_2$.
In order to confirm this hypothesis, we perform first principles calculations for this MLC as a function of $x$ (see Fig. 8 in SI ).
We find that the La-Pb alloying allows a perfect control of the doping level due to the large work function difference between LaSe and SnSe$_2$ and an insulator-to-metal transition  occurs in SnSe$_2$.
At $x=1.00$ the Fermi level perfectly matches the inflection point, as shown in  Fig. \ref{fig:5}.
It is worth noting that at this high La concentration, some LaSe bands cross the Fermi level close to the K point and along $\Gamma$K, however their contribution to the total density of states is marginal.
In Fig. \ref{fig:5} we also compare the MLC surface electronic structure with the one of an isolated layer (red line). As it can be seen, there is a substantial band distortion with respect to the isolated single layer. A better description of the surface electronic structure is obtained by replacing the LaSe layer with a uniformly positive charged potential barrier, as in a single gate field effect transistor setup by using the method developed in Ref. ~\cite{FET}. The electronic structure of an isolated SnSe$_2$ layer under this approximation is the green line in Fig. \ref{fig:5}, in perfect agreement with the complete calculation of the MLC surface electronic structure both for what concerns the band bending at the Fermi level (some deviations are seen in the empty states close to zone center) and for the position of the valence band top. We attribute the band-bending occurring at the K high-symmetry point to a modification
of the intralayer spacing between Sn and Se in SnSe$_{2}$ due to the charging of the monolayer (a table with intralayer spacing comparisons can be found in Fig. 7 of SI).

This result shows that it is possible via Pb/La alloying in the RS layers to set the Fermi level at the Van Hove singularity. Furthermore, it shows that the LaSe Q-layer can be assimilated to a capacitor plate in a Field  Effect Transistor (FET) (see Fig. \ref{fig:1}(a)). This remains true even for the SnSe$_2$ bilayers in the bulk of the sample, i.e. the full MLC can be assimilated to several field-effect transistors stacked periodically along the z-axis of the MLC, as shown in Fig. \ref{fig:1}.

As superconductivity is a bulk property, we must simulate the complete 3D crystal. The calculation of the vibrational properties and electron-phonon coupling for the complete MLC is, however, a very cumbersome task due to the large number of atoms. We then proceed differently, namely we consider a SnSe$_2$ bilayer in a field effect configuration as in Fig. \ref{fig:1} with a $+0.7$ charge on each of the two plates (double gate configuration). In order to prevent the ions from moving too close to the gate electrodes, a potential barrier is placed before the gates, and the total charge of the system is maintained equal to zero~\cite{FET}. Additional details on these calculations can be found in the SI. 
We have verified that this approach gives geometries for the SnSe$_2$ bilayer in excellent agreement with the complete MLC structural optimization. Furthermore the electronic density of states of the MLC and that of the monolayer in double gate configuration are practically indistinguishable, as shown in Fig. \ref{fig:5}.

We then calculate the phonon dispersion ($\omega_{\mathbf{q}\nu}$) and the electron-phonon coupling $\lambda_{\mathbf{q}\nu}$ for each mode $\nu$ of phonon crystal momentum $\mathbf{q}$ in double gate geometry. From these quantities we obtain the Eliashberg function $\alpha^2F(\omega)=\frac{1}{2 N_q}\sum_{\mathbf{q}\nu} \lambda_{\mathbf{q}\nu}\omega_{\mathbf{q}\nu}\delta(\omega-\omega_{\mathbf{q}\nu})$ and the
average electron-phonon coupling $\lambda=\frac{1}{N_q}\sum_{\mathbf{q}\nu}\lambda_{\mathbf{q}\nu}=0.6$, $N_q$ being the number of points in the phonon momentum grid used to calculate the average (we used a 96$\times$96$\times$1 $\mathbf{q}$-grid, see the SI).
These quantities are plotted in Fig. \ref{fig:5} (b).
Approximately $30\%$ of the coupling arises from the Einstein optical modes
at $\approx 45-50$ cm$^{-1}$, while the rest of the coupling is uniformly distributed throughout the other modes. The phonon density of states (not shown) is very similar to the Eliashberg function.

We calculate the superconducting critical temperature
 by  solving the anisotropic Migdal-Eliashberg equations~\cite{MigdalEliashberg}, as implemented in the EPIq software ~\cite{Marini_2023,calandraprofetamauri}, and by assuming $\mu^{*} = 0.1$, obtaining a superconducting critical temperature of $T_{c}=3.5$ K (see SI for details on Migdal-Eliashberg calculations). This result matches well with the $T_c=4.8$ K detected in ultrathin Li-intercalated SnSe$_2$ via field effect gating and demonstrates that superconductivity can emerge in MLC from pristine components that are not superconducting.

In conclusion, by performing extensive first principles electronic structure calculations on misfit layer compounds we unveiled the mechanism ruling charge transfer in these systems. In particular, due to their large work functions, we showed that TMDs are always acceptors while rocksalts are always donors. The electron density  that can be injected in the TMD layers can be as high as 6$\times$10$^{14}$~e$^-$cm$^{-2}$, sensibly larger than in ordinary field-effect transistors.

We have shown that the charging of the TMD layers can be efficiently controlled via the La$\leftrightarrow$Pb substitution. Most interesting, by replacing each RS Q-layer with a charged plate and a barrier, we have shown that the surface of the MLC behaves as a single gated field-effect transistor while the bulk can be seen as a periodic arrangement of double-gated field effect transistor.

Finally and most important, we have shown that from the knowledge of the RS and TMD constituents it is possible to infer the amount of charge transfer to the TMD layers in the MLC and to predict the physical properties of the heterostructure. As a practical demonstration, we showed that emergent superconductivity occurs in (LaSe)$_{1.27}$(SnSe2)$_2$ via a 1.4 eV Fermi level shift induced by the presence RS Q-layers in the misfit.
The methodology developed in this work paves the way to the synthesis and design of misfit compounds with
tailored physical properties.

\section{Acknowledgements}
We acknowledge EuroHPC for awarding us access to the LUMI supercomputer (grant number $465000468$). We acknowledge support from the European Union's Horizon 2020 research and innovation programme Graphene Flagship under grant agreement No $881603$. 

\section{Supporting Information}
Contains:
\begin{itemize}
  \item I. Geometrical Details of MLCs.
  \item II. Technical details.
  \item III. Band Alignment Calculation.
  \item IV. Band unfolding method applied to (La$_{x}$Pb$_{1-x}$Se)$_{1.18}$(TiSe$_{2}$)$_{2}$.
  \item V. Doping-induced Superconductivity.
\end{itemize} 

\bibliography{bibliography}
\providecommand{\latin}[1]{#1}
\makeatletter
\providecommand{\doi}
  {\begingroup\let\do\@makeother\dospecials
  \catcode`\{=1 \catcode`\}=2 \doi@aux}
\providecommand{\doi@aux}[1]{\endgroup\texttt{#1}}
\makeatother
\providecommand*\mcitethebibliography{\thebibliography}
\csname @ifundefined\endcsname{endmcitethebibliography}
  {\let\endmcitethebibliography\endthebibliography}{}

\end{document}